\documentstyle[preprint,aps,epsfig]{revtex}
\tightenlines
\begin{document}

\title{Arbitrary Force-Constant Changes in the Crystal Impurity
Problem}

\author{{Philip D. Mannheim\footnote{Email address: 
philip.mannheim@uconn.edu}} \\
\normalsize{Department of Physics,
University of Connecticut, Storrs, CT 06269}\\
}

\date{December 8, 2005}

\maketitle

\begin{abstract}

In a previous study of the dynamics of crystals with substitutional
point defects we had obtained simple and exact expressions for the
positions of the perturbed crystal modes and the intensities in them in
the case in which both the host-host and the host-impurity
force-constants were taken to be central and nearest neighbor. Such
expressions required knowledge of only one pure crystal lattice Green's
function, the one at the defect site itself. In this paper we extend our
previous study to incorporate non-central force-constants as well. We
find that the same simple expressions which we had previously obtained in
the nearest neighbor central force-constant case also hold without any
modification at all in the case of isotropic force-constant changes in
crystals whose host-host force-constants are isotropic, and in the
particular mixed case in which the fractional changes in the central and
isotropic components of the force-constants are equal to each other. In
the most general arbitrary nearest neighbor force-constant case we obtain
a reasonably compact exact expression for the dynamics of the perturbed
modes which only involves a total of two pure crystal lattice Green's
functions.

\end{abstract}

\section{Introduction}

Studies of nuclear resonant inelastic x-ray scattering made possible by
the advent of dedicated synchrotron rings have generated renewed interest
in the crystal impurity problem (see e.g.
\cite{Seto2000,Hu2003,Parlinski2004}). In the typical experimental set-up
x-rays are inelastically scattered off $\rm{M\ddot{o}ssbauer}$ active
nuclei embedded in host materials such as crystals. Such
$\rm{M\ddot{o}ssbauer}$ active nuclei act as impurities in the otherwise
perfect host crystals into which they are inserted, and lead to a
modification of the crystal dynamics of the host, with it being the
response of the perturbed system rather than that of the host itself
which is then measured in the inelastic x-ray scattering process. The
general theory for such impurity-induced modifications can for instance
be found in \cite{Maradudin1965} where exact, analytic expressions for
the effect of a mass change at the defect site are given. For the case of
defect-induced force-constant changes equally exact, analytic
expressions have been given \cite{Mannheim1968,Mannheim1971} in the
particular case where both the host-host and host-defect force-constants
are central and nearest neighbor. Specifically, for such a situation it
was found in the physically interesting  central force-constant
body-centered and face-centered cubic crystal host cases that for both of
them the positions of the frequencies of the impure crystal were given as
the solutions to  
\begin{equation}
1-\rho(\omega^{2})S(\omega^{2})=0~~,
\label{1}
\end{equation}
where $\rho(\omega^{2})$ is given by
\begin{equation}
\rho(\omega^{ 2})=\frac{M}{M^{\prime}}-1 +\frac{M\omega^{
2}}{A_{xx}(0,0)}\left(1-\frac{A_{xx}(0,0)}{A^{\prime}_{xx}(0,0)}\right)
=\frac{M}{M^{\prime}}-1 +\frac{2\omega^{
2}}{\omega_{\rm
max}^2}\left(1-\frac{A_{xx}(0,0)}{A^{\prime}_{xx}(0,0)}\right)~~,
\label{2}
\end{equation}
$M$ and $M^{\prime}$ are the host and defect atom masses, 
$A_{xx}(0,0)$ and $A^{\prime}_{xx}(0,0)$ are the pure and impure self
force-constants at the defect site, and the function $S(\omega^{2})$ is
given by
\begin{equation}
S(\omega^{ 2})=-1-M\omega^{2}g_0
=
\int_0^{\omega^2_{\rm max}}d\omega^{\prime 2}\frac{\omega^{\prime
2}\nu(\omega^{\prime 2})}{(\omega^{ 2}-\omega^{\prime 2})}  
\label{3}
\end{equation}
as integrated over the density of squared eigenfrequencies of the pure
crystal \cite{footnote1}. Additionally, the amplitude of vibration of the
defect as defined via the impure crystal defect displacement normal mode
expansion
$u_{\alpha}(t)=(\hbar/2\omega)^{1/2}\sum_{\omega}
\chi_{\alpha}(0,\omega^2)[ a_{\omega}e^{-i\omega
t}+a^{\dagger}_{\omega}e^{i\omega t}]$ ($\alpha=x,y,z$) was found to be
given by
\begin{equation}
|\chi^2(0,\omega^2)|=\sum
_{\alpha}|\chi_{\alpha}^2(0,\omega^2)|
=\frac{1}{MN}\left(\frac{M}{M^{\prime}}\right)^2
\bigg{[}\frac{1}{[1-\rho(\omega^{2})S_P(\omega^{2})]^2+ [\pi
\omega^2\nu(\omega^2)\rho(\omega^2)]^2}\bigg{]}
\label{4}
\end{equation}
for in-band modes with $\omega^2 \leq \omega_{\rm max}^2$
($S_P(\omega^{2})$ is the principal value of $S(\omega^{2})$), while
being given by
\begin{equation}
|\chi^2(0,\omega_L^2)|=\frac{1}{M}\left(\frac{M}{M^{\prime}}\right)^2
\bigg{[}\frac{1}{\rho^2(\omega_L^{2})T(\omega_L^{2})+
M/M^{\prime}-[1+\rho(\omega_L^2)]^2}\bigg{]}
\label{5}
\end{equation}
for localized modes which obey $1-\rho(\omega_L^{2})S(\omega_L^{2})=0$
with frequencies $\omega_L^2$ outside the band, with the function
$T(\omega_L^2)$ being given by 
\begin{equation}
T(\omega_L^{ 2})=\omega_L^4
\int_0^{\omega^2_{\rm max}}d\omega^{\prime 2}\frac{\nu(\omega^{\prime
2})}{(\omega_L^{ 2}-\omega^{\prime 2})^2}~~. 
\label{6}
\end{equation}
For inelastic $\rm{M\ddot{o}ssbauer}$ studies it is the quantity
$|\chi^2(0,\omega^2)|$ which then gives the probability for nuclear
resonant inelastic x-ray scattering at frequency $\omega$, with the
experimentally measurable PDOS (the so-called partial density of states)
at energy
$E$ discussed for instance in \cite{Hu2003} being given as
$D(E)=M^{\prime}|\chi^2(0,\omega^2)|\nu(\omega)$ \cite{footnote2}. In the
present paper we extend the analysis of
\cite{Mannheim1968,Mannheim1971} to incorporate non-central nearest
neighbor force-constants as well, an occurrence which is also of
experimental concern.

\section{Setting up the Force-Constant Change Problem}

In the standard harmonic approximation the equations of motion for the
displacements from equilibrium $e^{-i\omega t}u_{\alpha}(\ell)$ of the
atoms of a pure 3N degree of freedom crystal lattice are given by
\begin{equation}
\sum_{\beta,\ell^{\prime}}\left[A_{\alpha \beta}(\ell, \ell^{\prime})
-w^2M(\ell^{\prime})\delta_{\alpha\beta}
\delta(\ell,\ell^{\prime})\right]u_{\beta}(\ell^{\prime})=0~~,
\label{7}
\end{equation}
where $\ell$ ranges from $0$ to $N-1$, $\alpha=x,y,z$, $M(\ell)=M$ 
is the mass of the atom at site $\ell$, and
$A_{\alpha\beta}(\ell,\ell^{\prime})$ are the pure crystal
force-constants. Similarly, for a system with a substitutional point
impurity of mass $M^{\prime}$ located at the origin of coordinates and
changed force-constants $A^{\prime}_{\alpha\beta}(\ell,\ell^{\prime})$,
the displacements from equilibrium are given as
$e^{-i\omega^{\prime} t}u_{\alpha}(\ell)$,  with Eq. (\ref{7}) 
being replaced by
\begin{equation}
\sum_{\beta,\ell^{\prime}}\left[A_{\alpha \beta}(\ell, \ell^{\prime})
-w^{\prime 2}M\delta_{\alpha\beta}
\delta(\ell,\ell^{\prime})\right]u_{\beta}(\ell^{\prime})
=\sum_{\beta,\ell^{\prime}}V_{\alpha \beta}(\ell,
\ell^{\prime})u_{\beta}(\ell^{\prime})~~,
\label{8}
\end{equation}
with the changes from the pure crystal case having been isolated in the
perturbation
\begin{equation}
V_{\alpha \beta}(\ell, \ell^{\prime})
=-w^{\prime 2}(M-M^{\prime})\delta_{\alpha\beta}
\delta(\ell,0)\delta(\ell^{\prime},0)+A_{\alpha \beta}(\ell, 
\ell^{\prime})-
A^{\prime}_{\alpha \beta}(\ell, \ell^{\prime})~~.
\label{9}
\end{equation}
A formal solution for the positions of the frequency modes which satisfy
Eq. (\ref{8}) can be obtained in terms of the pure crystal lattice
Green's functions as evaluated in the lattice site representation.
Specifically, one first introduces the dynamical matrix of the pure
crystal 
\begin{equation}
D_{\alpha\beta}(\vec{\bf
k})=\frac{1}{M}\sum_{\ell}A_{\alpha\beta}(0,\ell)e^{-i\vec{\bf k} \cdot
\vec{\bf R}(\ell)}
\label{10}
\end{equation}
as expressed in terms of the phonon modes $\vec{\bf k}$ of the
translational invariant pure crystal, and then defines its eigenvectors
and eigenvalues according to
\begin{eqnarray}
\sum_{\beta}D_{\alpha\beta}(\vec{\bf
k})\sigma_{\beta}^{j}(\vec{\bf
k})&=&\omega_{j}^2(\vec{\bf k})\sigma_{\alpha}^{j}(\vec{\bf k})~~,
\nonumber \\
\sum_{\alpha}\sigma_{\alpha}^{*j}(\vec{\bf
k})\sigma_{\alpha}^{j^{\prime}}(\vec{\bf k})&=&\delta_{jj^{\prime}}~~,~~
\sum_{j}\sigma_{\alpha}^{*j}(\vec{\bf
k})\sigma_{\beta}^{j}(\vec{\bf k})=\delta_{\alpha\beta}~~,
\label{11}
\end{eqnarray}
and uses them to construct the pure
crystal lattice Green's functions according to
\begin{equation}
g_{\alpha\beta}(\omega;\ell,\ell^{\prime})=
\frac{1}{NM}\sum_{\vec{\bf k},j}\frac{\sigma_{\alpha}^{*j}
(\vec{\bf k})\sigma_{\beta}^{j}(\vec{\bf k})
e^{i\vec{\bf k}\cdot[\vec{\bf R}(\ell^{\prime})-\vec{\bf
R}(\ell)]}}{[\omega_{j}^2(\vec{\bf k})-\omega^2]}
\label{12}
\end{equation}
as summed over the three polarizations $j=(1,2,3)$ and $N$ momentum
vectors $\vec{\bf k}$ of all the modes in the Brillouin zone. As
constructed these Green's functions obey
\begin{equation}
\sum_{\ell,\beta}A_{\alpha\beta}(0,\ell)g_{\alpha^{\prime}\beta}
(\omega;\ell,\ell^{\prime})=
M\omega^2g_{\alpha^{\prime}\alpha}(\omega;0,\ell^{\prime})
+\frac{\delta_{\alpha,\alpha^{\prime}}}{N}\sum_{\vec{\bf k}}
e^{i\vec{\bf k}\cdot\vec{\bf R}(\ell^{\prime})}~~,
\label{13}
\end{equation}
and thus immediately allow us to solve Eq. (\ref{8}) in the form
\begin{equation}
u_{\alpha}(\ell)=\sum_{\ell^{\prime},\ell^{\prime\prime},
\beta,\gamma}g_{\alpha\beta} (\omega^{\prime};\ell,\ell^{\prime})
V_{\beta\gamma}(\ell^{\prime},\ell^{\prime\prime})
u_{\gamma}(\ell^{\prime\prime})~~,
\label{14}
\end{equation}
with the eigenmodes then being given as the solutions to the
($3N$-dimensional) determinantal condition 
\begin{equation}
|1-G_0V|=0~~,
\label{15}
\end{equation}
as written in an obvious notation.

For an explicit determination of the defect amplitude of vibration
$|\chi^2(0,\omega^2)|$ in any given case of interest, one needs to
introduce the lattice Green's functions of the impure crystal ($G$) which
are related to the above pure crystal lattice Green's functions ($G_0$)
via 
\begin{equation}
G=G_0+G_0VG_0+G_0VG_0VG_0+...=(1-G_0V)^{-1}G_0~~.
\label{16}
\end{equation}
As already noted above, the frequencies of the perturbed modes are given
as the solutions to $|1-G_0V|=0$,
while the intensities of interest in the impure and pure modes at the
defect site (respectively $|\chi^2(0,\omega^2)|$ and
$|\chi^2_{\rm pure}(0,\omega^2)|=1/NM$) are related by
(see e.g. \cite{Mannheim1971})
\begin{equation}
{\rm Im}G_{xx}(0,0)=\frac{|\chi^2(0,\omega^2)|}{|\chi^2_{\rm
pure}(0,\omega^2)|}{\rm Im}(G_0)_{xx}(0,0)=|\chi^2(0,\omega^2)|\pi
N\nu(\omega^2)~~,
\label{17}
\end{equation}
where $\nu(\omega^2)$ is the density of squared frequency states
of the pure crystal as normalized to one. It is the quantity ${\rm
Im}G_{xx}(0,0)$ which determines the response of the system to an
external probe, with the PDOS which gives the probability for
nuclear resonant inelastic x-ray scattering at frequency $\omega$ being
given by $D(E)=M^{\prime}|\chi^2(0,\omega^2)|\nu(\omega)=2\omega
M^{\prime}|\chi^2(0,\omega^2)|\nu(\omega^2)=(2\omega
M^{\prime} /\pi N){\rm Im}G_{xx}(0,0)$. The defect amplitude
$|\chi^2(0,\omega^2)|$ modulates the response of the system and it is
thus its determination which is needed for inelastic
$\rm{M\ddot{o}ssbauer}$ studies \cite{footnote3}.

In an actual application of Eq. (\ref{17}) to determine the needed
$|\chi^2(0,\omega^2)|$, the key step is in inverting the matrix
$1-G_0V$ as needed to obtain $G_{xx}(0,0)$ via Eq. (\ref{16}). And even if
one restricts to nearest neighbor force-constants only (which we will in
fact do here and throughout) and to just a single substitutional defect,
in the force-constant change case the matrix $V_{\alpha \beta}(\ell,
\ell^{\prime})$ will involve the defect and every single one of its
host nearest neighbor atoms. For a simple cubic crystal for instance this
defect-nearest neighbor complex has 21 degrees of freedom (seven
3-dimensional vibrations due to one defect and six neighbors), while for
the body-centered and face-centered cubic crystals the complexes
respectively have 27 and 39 relevant degrees of freedom, to initially
make the matrix $V_{\alpha \beta}(\ell,\ell^{\prime})$ and the
non-trivial sector of the matrix $1-G_0V$ quite large.  However, because of
the high $O_h$ symmetry at the defect site the
$V_{\alpha\beta}(\ell,\ell^{\prime})$ matrix can be block diagonalized in
the irreducible representations of the octahedral group, with 
the relevant decompositions in the simple, body-centered and
face-centered cubic crystals being of the form (see e.g.
\cite{Maradudin1965})
\begin{eqnarray}
\Gamma_{\rm sc}&=&A_{1g}+E_g+F_{1g}+
F_{2g}+3F_{1u}+F_{2u}
\nonumber \\
\Gamma_{\rm bcc}&=&A_{1g}+E_g+F_{1g}+
2F_{2g}+A_{2u}+E_u+3F_{1u}+F_{2u}
\nonumber \\
\Gamma_{\rm fcc}&=&A_{1g}+A_{2g}+2E_g+2F_{1g}+
2F_{2g}+A_{2u}+E_u+4F_{1u}+2F_{2u}~~.
\label{18}
\end{eqnarray}
Then, since the displacement of the defect atom
itself transforms as a 3-dimensional vector, the defect displacements
must be located entirely within the $F_{1u}$ modes, with all of the other
irreducible representations being built out of displacements of
appropriate linear combinations which involve the nearest neighbors of the
defect alone. Since it is the defect response to which external probes
such as nuclear resonant inelastic x-ray scattering couple, the sector of
$V_{\alpha\beta}(\ell,\ell^{\prime})$ which is relevant for such
scattering thus reduces to respective 3-dimensional, 3-dimensional and
4-dimensional blocks in Eq. (\ref{16}) each one of which is
itself threefold degenerate. For the physically interesting body-centered
cubic crystal the normalized 3-dimensional $3F_{1u}$ mode basis is given
(in body-centered cubic crystal site notation) by 
\begin{eqnarray}
\alpha_0&=&u_x(0,0,0)~~,
\nonumber \\
\alpha_1&=&\frac{1}{2\sqrt{2}}\bigg{[}u_x(1,1,1)
+u_x(\bar{1},1,1)+u_x(\bar{1},\bar{1},1)+u_x(1,\bar{1},1)
\nonumber \\
&&+u_x(\bar{1},\bar{1},\bar{1})+u_x(1,\bar{1},\bar{1})
+u_x(1,1,\bar{1})+u_x(\bar{1},1,\bar{1})\bigg{]}~~,
\nonumber \\
\alpha_2&=&\frac{1}{4}\bigg{[}
u_y(1,1,1)+u_z(1,1,1)
-u_y(\bar{1},1,1)-u_z(\bar{1},1,1)
\nonumber \\
&&+u_y(\bar{1},\bar{1},1)-u_z(\bar{1},\bar{1},1)
-u_y(1,\bar{1},1)+u_z(1,\bar{1},1)
\nonumber \\
&&+u_y(\bar{1},\bar{1},\bar{1})+u_z(\bar{1},\bar{1},\bar{1})
-u_y(1,\bar{1},\bar{1})-u_z(1,\bar{1},\bar{1})
\nonumber \\
&&+u_y(1,1,\bar{1})-u_z(1,1,\bar{1})
-u_y(\bar{1},1,\bar{1})+u_z(\bar{1},1,\bar{1})
\bigg{]}~~, 
\label{19}
\end{eqnarray}
while for the equally interesting face-centered cubic crystal the
normalized 4-dimensional $4F_{1u}$ mode basis is given (in face-centered
cubic crystal site notation) by
\begin{eqnarray}
\alpha_0&=&u_x(0,0,0)~~,
\nonumber \\
\alpha_1&=&\frac{1}{2\sqrt{2}}\bigg{[}u_x(1,1,0)
+u_x(\bar{1},\bar{1},0)+u_x(1,0,1)+u_x(\bar{1},0,\bar{1})
\nonumber \\
&&+u_x(1,\bar{1},0)+u_x(\bar{1},1,0)
+u_x(\bar{1},0,1)+u_x(1,0,\bar{1})\bigg{]}~~,
\nonumber \\
\alpha_2&=&\frac{1}{2\sqrt{2}}\bigg{[}u_y(1,1,0)+u_y(\bar{1},\bar{1},0)
+u_z(1,0,1)+u_z(\bar{1},0,\bar{1})
\nonumber \\
&&-u_y(1,\bar{1},0)-u_y(\bar{1},1,0)
-u_z(\bar{1},0,1)-u_z(1,0,\bar{1})\bigg{]}~~,
\nonumber \\
\alpha_3&=&\frac{1}{2}\bigg{[}u_x(0,1,1)+u_x(0,\bar{1},\bar{1})
+u_x(0,1,\bar{1})
+u_x(0,\bar{1},1)\bigg{]}~~. 
\label{20}
\end{eqnarray}
Degenerate with each of these bases are two others, one based on
$u_y(0,0,0)$ and the other on $u_z(0,0,0)$. With the body-centered and
face-centered cubic crystals being each other's reciprocal lattice, in
the harmonic approximation where momentum and position are treated
equivalently, the defect responses in the two cases must be identical
\cite{Mannheim1968}, and it will thus suffice in the following to treat
just one of the two of them. And with the body-centered cubic basis of Eq.
(\ref{19}) having lower dimensionality than the face-centered cubic
crystal basis given in Eq. (\ref{20}), in the following we shall treat the
body-centered cubic alone. (While we do not treat the simple cubic
crystal case here, for it one can anticipate results analogous to those
we shall provide below for the body-centered cubic case.) 

In terms of the basis of Eq. (\ref{19}) the matrix of the pure
crystal $G_0$ in the body-centered cubic $F_{1u}$ mode can be
written very compactly as \cite{Mannheim1971}
\begin{eqnarray}
[G_0]_{F_{1u}} = \pmatrix{
g_0&2\surd{2}g_1&4g_2\cr
2\surd{2}g_1&Q&\surd{2}R \cr
4g_2&\surd{2}R&S+T  \cr}~~,
\label{21}
\end{eqnarray}
where we have introduced the notation 
\begin{eqnarray}
g_0&=&g_{xx}(000)~~,~~g_1=g_{xx}(111)~~,~~g_2=g_{xy}(111)~~,~~R=g_{xy}(222)+g_{xy}(220)~~,
\nonumber \\
Q&=&g_0+g_{xx}(222)
+g_{xx}(200)+g_{xx}(022)+2g_{xx}(220)+2g_{xx}(020)~~,
\nonumber \\
S&=&g_0+g_{yy}(222)
-g_{yy}(020)-g_{yy}(202)~~,~~T=g_{yz}(222)-g_{yz}(022)
\label{22}
\end{eqnarray}
which takes advantage of the translational invariance of the pure
crystal lattice to set
$g_{\alpha\beta}(\omega;\ell,\ell^{\prime})
=g_{\alpha\beta}(\omega;\ell-\ell^{\prime},0)
= g_{\alpha\beta}(\ell-\ell^{\prime})$. The pure
crystal Green's functions which appear in Eq. (\ref{21}) are not
completely independent of each other, as some of them are related via the
general Eq. (\ref{13}). Specifically, if Eq. (\ref{13}) is restricted to
nearest neighbor force-constants, for the body-centered cubic we obtain
the relations \cite{Mannheim1971}  
\begin{eqnarray}
&&A_{xx}(0,0)g_0+8A_{xx}(111)g_1+16A_{xy}(111)g_2=1+M\omega^2 g_0~~,
\nonumber \\
&&8A_{xy}(111)g_2=0~~,
\nonumber \\
&&A_{xx}(0,0)g_1+A_{xx}(111)Q+2A_{xy}(111)R=M\omega^2 g_1~,
\nonumber \\
&&A_{xx}(0,0)g_2+A_{xx}(111)R+A_{xy}(111)(S+T)=M\omega^2 g_2~~.
\label{23}
\end{eqnarray}
(Our notation here is to set $A_{\alpha\beta}(\ell,\ell^{\prime})
=A_{\alpha\beta}(\ell-\ell^{\prime},0)
= A_{\alpha\beta}(\ell-\ell^{\prime})$, but to use $A_{\alpha\beta}(0,0)$
to denote $A_{\alpha\beta}(\ell=0,\ell^{\prime}=0)$.)
In establishing Eq. (\ref{23}) we have used the fact that for the
body-centered cubic crystal the lattice delta function
$(1/N)\sum_{\vec{\bf k}} e^{i\vec{\bf k}\cdot\vec{\bf
R}(\ell^{\prime})}$ with $\vec{\bf R}(\ell^{\prime})=(p,q,r)$ has the
property that it is equal to one if
$p+q+r$ is an even integer and equal to zero otherwise \cite{footnote4}.
As we see, via the use of Eq. (\ref{23}) only two of the Green's
functions which appear in Eq. (\ref{21}) are independent. Evaluation of
Eq. (\ref{16}) can thus involve no more than two independent pure lattice
Green's functions.

In order to be able characterize the various pure crystal force-constants
$A_{\alpha\beta}(\ell,\ell^{\prime})$ which appear in Eq. (\ref{23}) as
well as their impure $A^{\prime}_{\alpha\beta}(\ell,\ell^{\prime})$
counterparts, we recall that in terms of the two-body host-host
interatomic potential $\phi(r)$, the force-constants between a host atom
vibrating around site $R_{\alpha}(\ell)$ and one vibrating around the
origin are defined as
\begin{equation}
A_{\alpha\beta}(\ell,0)=-
\left(\frac{\partial^2\phi(r)}{\partial
u_{\alpha}(\ell)\partial u_{\beta}(\ell)}\right)\bigg{|}_0=
-\left(\frac{\phi^{\prime\prime}(r)}{r^2}
-\frac{\phi^{\prime}(r)}{r^3}\right)R_{\alpha}(\ell)R_{\beta}(\ell)
-\frac{\phi^{\prime}(r)}{r}\delta_{\alpha\beta}~~,
\label{24}
\end{equation}
as calculated at the equilibrium lattice separation between the host
atoms. Force-constants for which $\phi^{\prime}(r)$ just happens to
vanish when the atoms are in their equilibrium positions are referred to
as being central, while those associated with potentials which obey
$\phi^{\prime}(r)/r=\phi^{\prime\prime}(r)$ at equilibrium are referred
to as being isotropic. In terms of the various $A_{\alpha\beta}(\ell,0)$
with $\ell \neq 0$ the self-force-constant at the origin is then given
via Newton's third law as the summation 
\begin{equation}
A_{\alpha\beta}(0,0)=-\sum_{\ell \neq
0}A_{\alpha\beta}(0,\ell)~~.
\label{25}
\end{equation}
For a nearest neighbor pure crystal the force-constants can thus be
characterized in terms of two parameters, viz.
\begin{equation}
\alpha=-\frac{1}{3}\left(\phi^{\prime\prime}(r)
-\frac{\phi^{\prime}(r)}{r}\right)~~,~~\beta
=-\frac{\phi^{\prime}(r)}{r}~~,
\label{26}
\end{equation}
in terms of which we obtain
\begin{equation}
A_{\alpha\beta}(111)= \pmatrix{
\alpha+\beta&\alpha&\alpha\cr
\alpha&\alpha+\beta&\alpha\cr
\alpha&\alpha&\alpha+\beta\cr}~~,~~
A_{\alpha\beta}(\bar{1}11)= \pmatrix{
\alpha+\beta&-\alpha&-\alpha\cr
-\alpha&\alpha+\beta&\alpha\cr
-\alpha&\alpha&\alpha+\beta\cr}
\label{27}
\end{equation}
and 
\begin{equation}
A_{xx}(0,0)=-8(\alpha+\beta)~~,~~A_{xy}(0,0)=0~~.
\label{28}
\end{equation}
Given Eqs. (\ref{27}) and (\ref{28}) we can now rewrite the Green's
function relations of Eq. (\ref{23}) in the convenient form
\begin{eqnarray}
8(\alpha+\beta)g_1&=&8(\alpha+\beta)g_0+1+M\omega^2 g_0~~,
\nonumber \\
g_2&=&0~~,
\nonumber \\
(\alpha+\beta)Q&=&M\omega^2 g_1+8(\alpha+\beta)g_1-2\alpha R~~,
\nonumber \\
\alpha(S+T)&=&-(\alpha+\beta)R~~.
\label{29}
\end{eqnarray}

When a defect is introduced substitutionally at a lattice site, the
interatomic potential $\hat{\phi}(r)$ between it and a host neighbor will
in principle differ from that between pairs of host atoms, leading to
modified force-constants 
\begin{equation}
A^{\prime}_{\alpha\beta}(\ell,0)=-
\left(\frac{\partial^2\hat{\phi}(r)}{\partial
u_{\alpha}(\ell)\partial u_{\beta}(\ell)}\right)\bigg{|}_0=
-\left(\frac{\hat{\phi}^{\prime\prime}(r)}{r^2}
-\frac{\hat{\phi}^{\prime}(r)}{r^3}\right)R_{\alpha}(\ell)R_{\beta}(\ell)
-\frac{\hat{\phi}^{\prime}(r)}{r}\delta_{\alpha\beta}~~,
\label{30}
\end{equation}
and
\begin{equation}
A^{\prime}_{\alpha\beta}(0,0)=-\sum_{\ell \neq
0}A^{\prime}_{\alpha\beta}(0,\ell)~~,
\label{31}
\end{equation}
as again calculated at the host atom lattice equilibrium separation (the
defect being inserted substitutionally), with the modified
force-constants then being characterized by two parameters
\begin{equation}
\hat{\alpha}=-\frac{1}{3}\left(\hat{\phi}^{\prime\prime}(r)
-\frac{\hat{\phi}^{\prime}(r)}{r}\right)~~,~~\hat{\beta}
=-\frac{\hat{\phi}^{\prime}(r)}{r}~~,
\label{32}
\end{equation}
whose relation to $\alpha$ and $\beta$ can be arbitrary \cite{footnote5}.
For the impure system we thus obtain 
\begin{equation}
A^{\prime}_{\alpha\beta}(111)= \pmatrix{
\hat{\alpha}+\hat{\beta}&\hat{\alpha}&\hat{\alpha}\cr
\hat{\alpha}&\hat{\alpha}+\hat{\beta}&\hat{\alpha}\cr
\hat{\alpha}&\hat{\alpha}&\hat{\alpha}+\hat{\beta}\cr
},~~
A^{\prime}_{\alpha\beta}(\bar{1}11)= \pmatrix{
\hat{\alpha}+\hat{\beta}&-\hat{\alpha}&-\hat{\alpha}\cr
-\hat{\alpha}&\hat{\alpha}+\hat{\beta}&\hat{\alpha}\cr
-\hat{\alpha}&\hat{\alpha}&\hat{\alpha}+\hat{\beta}\cr}
\label{33}
\end{equation}
and 
\begin{equation}
A^{\prime}_{xx}(0,0)=-8(\hat{\alpha}+\hat{\beta})~~,~~
A^{\prime}_{xy}(0,0)=0~~.
\label{34}
\end{equation}
Finally, on introducing the parameters
\begin{equation}
X=\alpha-\hat{\alpha}~~,~~
Y=\beta-\hat{\beta}~~.
\label{35}
\end{equation}
we can write the matrix elements of the perturbation
$V_{\alpha\beta}(\ell,\ell^{\prime})$ in the $F_{1u}$ mode as 
\begin{eqnarray}
[V]_{F_{1u}} = \pmatrix{
-\omega^2(M-M^{\prime})-8X-8Y&2\surd{2}(X+Y)&4X\cr
2\surd{2}(X+Y)&-(X+Y)&-\surd{2}X \cr
4X&-\surd{2}X&-2X-Y  \cr}~~.
\label{36}
\end{eqnarray}
Armed with Eqs. (\ref{21}), (\ref{29}) and (\ref{36}) we can now proceed
to an evaluation of $(1-G_0V)^{-1}G_0$.

\section{Solving the Force-Constant Change Problem}

To solve the problem we first proceed symbolically and set 
\begin{eqnarray}
[1-G_0V]_{F_{1u}} = \pmatrix{
a&b&c\cr
d&e&f\cr
g&h&i  \cr}~~,
\label{37}
\end{eqnarray}
where 
\begin{eqnarray}
a&=&1-\omega^2(M^{\prime}-M)g_0+8(X+Y)(g_0-g_1)~~,
\nonumber \\
b&=&-2\surd{2}(X+Y)(g_0-g_1)~~,
\nonumber \\
c&=&-4X(g_0-g_1)~~,
\nonumber \\
d&=&-2\surd{2}[\omega^2(M^{\prime}-M)g_1-(X+Y)(8g_1-Q)+2XR]~~,
\nonumber \\
e&=&1-(X+Y)(8g_1-Q)+2XR~~,
\nonumber \\
f&=&\surd{2}[-X(8g_1-Q)+(2X+Y)R]~~,
\nonumber \\
g&=&-4(X+Y)R-4X(S+T)~~,
\nonumber \\
h&=&\surd{2}[(X+Y)R+X(S+T)]~~,
\nonumber \\
i&=&1+2XR+(2X+Y)(S+T)~~.
\label{38}
\end{eqnarray}
In terms of these symbolic quantities the inverse matrix is given by 
\begin{eqnarray}
[1-G_0V]^{-1}_{F_{1u}} = \frac{1}{\Delta}\pmatrix{
ei-fh&hc-ib&fb-ce\cr
fg-di&ia-gc&cd-af\cr
dh-ge&gb-ha&ae-bd  \cr}~~,
\label{39}
\end{eqnarray}
where the determinant of $[1-G_0V]_{F_{1u}}$ is given by 
\begin{equation}
\Delta=a(ei-fh)+d(hc-ib)+g(bf-ce)~~.
\label{40}
\end{equation}
From Eqs. (\ref{16}) and (\ref{21}) and recalling the vanishing of $g_2$,
it then follows that the defect site component of the impure crystal
Green's function 
$G$ is given in closed form by
\begin{equation}
G_{xx}(0,0)=\frac{1}{\Delta}[(ei-fh)g_0+2\surd{2}(hc-ib)g_1]~~,
\label{41}
\end{equation}
with the matrix $1-G_0V$ having been inverted analytically.

For computational purposes we note that in both $\Delta$ and $G_{xx}(0,0)$
it is just three symbolic combinations, viz. $ei-fh$, $hc-ib$
and $bf-ce$, which are needed. From Eq. (\ref{38}) these combinations are
readily found to evaluate to 
\begin{eqnarray}
ei-fh&=&1+4XR+(2X+Y)(S+T)-(X+Y)(8g_1-Q)
\nonumber \\
&&+(3XY+Y^2)[-(S+T)(8g_1-Q)-2R^2]~~,
\nonumber \\
hc-ib&=&2\surd{2}(g_0-g_1)[X+Y+(3XY+Y^2)(S+T)]~~,
\nonumber \\
bf-ce&=&-4(g_0-g_1)[-X+(3XY+Y^2)R]~~.
\label{42}
\end{eqnarray}
Further algebra shows that the numerator in Eq. (\ref{41}) evaluates to
\begin{eqnarray}
{\rm NUM}&=&(ei-fh)g_0+2\surd{2}(hc-ib)g_1
\nonumber \\
&=&g_0[1+4XR+(2X+Y)(S+T)+(X+Y)Q]
\nonumber \\
&&-8(X+Y)g_1^2+(3XY+Y^2)[(S+T)(g_0Q-8g_1^2)-2g_0R^2]~~,
\label{43}
\end{eqnarray}
while the denominator evaluates to 
\begin{eqnarray}
\Delta&=&1+4XR+(X+Y)(8g_0-16g_1+Q)+(2X+Y)(S+T)
\nonumber \\
&&+(3XY+Y^2)[(S+T)(8g_0-16g_1+Q)-2R^2] -8(M-M^{\prime})\omega^2(X+Y)g_1^2
\nonumber \\
&&+(M-M^{\prime})\omega^2g_0[1+4XR+(2X+Y)(S+T)+(X+Y)Q]
\nonumber \\
&&+(M-M^{\prime})\omega^2(3XY+Y^2)[(S+T)(g_0Q-8g_1^2)-2g_0R^2]~~.
\label{44}
\end{eqnarray}

To simplify the problem further we now utilize the pure crystal Green's
function relations given in Eq. (\ref{29}), to find, following
a fair amount of algebra, that the numerator and denominator in
$G_{xx}(0,0)$ reduce to
\begin{eqnarray}
{\rm NUM}&=&\frac{g_0}{\alpha(\alpha+\beta)}
\left[\alpha(\alpha+\beta)
-2\beta^2XR-(3\alpha^2+2\alpha\beta+\beta^2)YR\right]
\nonumber \\
&&+\frac{g_1}{\alpha(\alpha+\beta)}\left[(\alpha+\beta)(3XY+Y^2)R
-\alpha(X+Y)\right]~~,
\label{45}
\end{eqnarray}
and
\begin{eqnarray}
\Delta&=&\frac{[1+(M-M^{\prime})\omega^2g_0]}{\alpha(\alpha+\beta)}
\left[\alpha(\alpha+\beta)
-2\beta^2XR-(3\alpha^2+2\alpha\beta+\beta^2)YR\right]
\nonumber \\
&&+\frac{[1+M\omega^2g_0-M^{\prime}\omega^2g_1]}{\alpha(\alpha+\beta)}
\left[(\alpha+\beta)(3XY+Y^2)R
-\alpha(X+Y)\right]~~,
\label{46}
\end{eqnarray}
where
\begin{equation}
g_1=g_0+\frac{[1+M\omega^2g_0]}{8(\alpha+\beta)}~~.
\label{47}
\end{equation}
In terms of the convenient functions
\begin{eqnarray}
\hat{R}&=&\frac{(\alpha+\beta)(3XY+Y^2)R}{\alpha(X+Y)}=
\frac{(\alpha+\beta)(\beta-\hat{\beta})(3\alpha-3\hat{\alpha}
+\beta-\hat{\beta})R}
{\alpha(\alpha-\hat{\alpha}+\beta-\hat{\beta})}~~,
\nonumber \\
\mu&=&\frac{2(\beta X-\alpha Y)^2}
{(\alpha+\beta)^2(3XY+Y^2)}
=\frac{2(\alpha\hat{\beta}-\beta\hat{\alpha})^2}
{(\alpha+\beta)^2
(\beta-\hat{\beta})(3\alpha-3\hat{\alpha}+\beta-\hat{\beta})}~~,
\label{48}
\end{eqnarray}
which obey
\begin{equation}
\mu\hat{R}=
\frac{2(\alpha\hat{\beta}-\beta\hat{\alpha})^2R}
{\alpha(\alpha+\beta)
(\alpha-\hat{\alpha}+\beta-\hat{\beta})}~~,
\label{49}
\end{equation}
the numerator and denominator in $G_{xx}(0,0)$ can be written more
compactly as
\begin{eqnarray}
{\rm NUM}&=&\frac{A^{\prime}_{xx}(0,0)}{A_{xx}(0,0)}
\left[g_0+\frac{(1+M\omega^2g_0)}{A_{xx}(0,0)}\left(\frac{
A_{xx}(0,0)}{A^{\prime}_{xx}(0,0)}-1\right)\right]
\left[1-\hat{R}\right]-\mu g_0\hat{R}
\nonumber \\
&=&\frac{1}{M\omega^2}\left[
\frac{A^{\prime}_{xx}(0,0)}{A_{xx}(0,0)}\left
(\rho(\omega^2)S(\omega^2)-\frac{M}{M^{\prime}}S(\omega^2)-1\right)
\left[1-\hat{R}\right]+\mu [S(\omega^2)+1]\hat{R}\right]~~,
\label{50}
\end{eqnarray}
and
\begin{eqnarray}
\Delta&=&\frac{A^{\prime}_{xx}(0,0)}{A_{xx}(0,0)}
\bigg{\{}1+M\omega^2g_0-M^{\prime}\omega^2g_0
+\frac{M^{\prime}\omega^2}{A_{xx}(0,0)}
-\frac{M^{\prime}\omega^2}{A^{\prime}_{xx}(0,0)}
\nonumber \\
&&
+\frac{MM^{\prime}\omega^4g_0}{A_{xx}(0,0)}
-\frac{MM^{\prime}\omega^4g_0}{A^{\prime}_{xx}(0,0)}
\bigg{\}}
\left[1-\hat{R}\right]
-\mu\hat{R}\left[1+(M-M^{\prime})\omega^2g_0\right]
\nonumber \\
&=&\frac{M^{\prime}}{M}\left\{\frac{A^{\prime}_{xx}(0,0)}{A_{xx}(0,0)}
\left[1-\rho(\omega^2)
S(\omega^2)\right]\left[1-\hat{R}\right]
-\mu\hat{R}\left[S(\omega^2)\left(1-\frac{M}{M^{\prime}}\right)
+1\right]\right\}
\label{51}
\end{eqnarray}
where $\rho(\omega^2)$ and $S(\omega^2)=-1-M\omega^2g_0$ were introduced
earlier in Eqs. (\ref{2}) and (\ref{3}). The form for $G_{xx}(0,0)={\rm
NUM}/\Delta$ implied by Eqs. (\ref{50}) and (\ref{51}) is our main
result, an exact relation for a body-centered cubic crystal under the
sole assumption of nearest neighbor force-constants, with there being no
restriction on the strengths of the central and non-central
force-constants in the pure crystal or the amount by which they might
change in the presence of the defect, with the evaluation of $G_{xx}(0,0)$
requiring a knowledge of only two pure crystal lattice Green's functions,
$g_0$ and $R$.

While the form of Eqs. (\ref{50}) and (\ref{51}) is completely general,
great simplification occurs whenever we can set $\mu\hat{R}=0$, i.e.
whenever the force-constant changes obey
\begin{equation}
\alpha\hat{\beta}-\beta\hat{\alpha}=0~~,
\label{52}
\end{equation}
since then $G_{xx}(0,0)$ reduces to 
\begin{equation}
G_{xx}(0,0)=\frac{\left[\rho(\omega^2)S(\omega^2)
-(M/M^{\prime})S(\omega^2) -1\right]}
{M^{\prime}\omega^2[1-\rho(\omega^2)S(\omega^2)]}~~,
\label{53}
\end{equation}
to then only depend on one pure lattice Green's function alone, the one
at the defect site itself. In this restricted case we see that the
positions of the eigenmodes are given as the solutions to 
\begin{equation}
1-\rho(\omega^2)S(\omega^2)=0~~,
\label{54}
\end{equation}
while on giving $\omega^2$ a small imaginary part in the complex plane and
recalling that 
\begin{equation}
\frac{1}{(\omega^2-\omega^{\prime
2}+i\epsilon)}=P\left(\frac{1}{\omega^2-\omega^{\prime
2}}\right) -i\pi\delta(\omega^2-\omega^{\prime 2})~~,
\label{55}
\end{equation}
find that the imaginary part of $G_{xx}(0,0)$ evaluates to
\begin{equation}
{\rm Im}G_{xx}(0,0)=
\frac{1}{M}\left(\frac{M}{M^{\prime}}\right)^2
\bigg{[}\frac{\pi\nu(\omega^2)}{[1-\rho(\omega^{2})S_P(\omega^{2})]^2+
[\pi
\omega^2\nu(\omega^2)\rho(\omega^2)]^2}\bigg{]}~~.
\label{56}
\end{equation}
With the imaginary part of the pure crystal $g_0$ evaluating to ${\rm
Im}g_0=\pi\nu(\omega^2)/M=N|\chi^2_{\rm
pure}(0,\omega^2)|\pi\nu(\omega^2)$, Eqs. (\ref{1}) and (\ref{4}) thus
follow \cite{footnote6}. With Eqs. (\ref{1}) and (\ref{4}) having
previously been obtained under the central force-constant restriction, we
see now that they in fact have far greater validity, since as well as the
pure central force-constant solution to Eq. (\ref{52}) in which
$\beta=\hat{\beta}=0$, Eq. (\ref{52}) also admits of other solutions.
Specifically it admits of a pure isotropic force-constant solution in
which $\alpha=\hat{\alpha}=0$, and also of solution in which the central
and isotropic components of the force-constants undergo the same
fractional change $\hat{\alpha}/\alpha =\hat{\beta}/\beta$.

While the family of solutions which obey
$\alpha\hat{\beta}-\beta\hat{\alpha}=0$ encompasses a large and
interesting class of force-constant changes, to go beyond this set of
solutions requires including the second Green's function $R$. Unlike the
Green's function $g_0$ whose evaluation depends only on the density of
states of the pure crystal according to
\begin{eqnarray}
g_0&=&g_{xx}(\omega;0,0)=
\frac{1}{3}[g_{xx}(\omega;0,0)+g_{yy}(\omega;0,0)+g_{zz}(\omega;0,0)]
\nonumber \\
&=&
\frac{1}{3NM}\sum_{\vec{\bf k},j}\frac{
[\sigma_{x}^{*j}(\vec{\bf k})\sigma_{x}^{j}(\vec{\bf k})
+\sigma_{y}^{*j}(\vec{\bf k})\sigma_{y}^{j}(\vec{\bf k})
+\sigma_{z}^{*j}(\vec{\bf k})\sigma_{z}^{j}(\vec{\bf k})]}
{[\omega_{j}^2(\vec{\bf k})-\omega^2]}
\nonumber \\
&=&\frac{1}{3NM}\sum_{\vec{\bf
k},j}\frac{1}{[\omega_{j}^2(\vec{\bf k})-\omega^2]}=
\frac{1}{M}\int_0^{\omega^2_{\rm
max}}d\omega^{\prime 2}\frac{\nu(\omega^{\prime 2})}{[\omega^{\prime
2}-\omega^2]}~~,
\label{57}
\end{eqnarray}
as can be seen from the definition of Eq. (\ref{12}), any other pure
lattice Green's function such as $R$ will involve a sum over momentum
modes which cannot be reduced to an equivalent sum over
eigenfrequencies. However, as noted in \cite{Mannheim1968} it is
possible to simply the evaluation of the lattice Green's functions
which involve a non-zero $\vec{\bf R}(\ell^{\prime})-\vec{\bf
R}(\ell)$ to some degree, as it is possible to perform the sum over
polarizations analytically. Specifically, it was shown that the general
$g_{\alpha\beta}(\omega;\ell,0)$ with $\ell \neq 0$ can be written in
closed form as 
\begin{equation}
g_{\alpha\beta}(\omega;\ell,0)=
-\frac{1}{NM}\sum_{\vec{\bf k}}\lambda_{\alpha\beta}(\vec{\bf k})
e^{-i\vec{\bf k}\cdot\vec{\bf R}(\ell)}
\label{58}
\end{equation}
where the different components of $\lambda_{\alpha\beta}(\vec{\bf k})$
can be written in terms of the pure crystal dynamical matrix
$D_{\alpha\beta}(\vec{\bf k})$ of Eq. (\ref{10}) as
\begin{equation}
\lambda_{xx}(\vec{\bf k})=
\frac{[\omega^4-\omega^2(D_{yy}+D_{zz})+D_{yy}D_{zz}-D_{yz}^2]}
{[\omega^6-\omega^4\sum D_{xx}+\omega^2(\sum D_{yy}D_{zz}-\sum
D_{xy}^2)-\prod D_{xx}+\sum D_{xx}D_{yz}^2-2\prod D_{xy}]}
\label{59}
\end{equation}
and 
\begin{equation}
\lambda_{xy}(\vec{\bf k})=
\frac{[\omega^2D_{xy}-D_{xy}D_{zz}+D_{xz}D_{yz}]}
{[\omega^6-\omega^4\sum D_{xx}+\omega^2(\sum D_{yy}D_{zz}-\sum
D_{xy}^2)-\prod D_{xx}+\sum D_{xx}D_{yz}^2-2\prod D_{xy}]}
\label{60}
\end{equation}
(the sums being taken cyclically), to thus reduce a determination of the
pure lattice Green's functions to a straightforward sum over the Brillouin
zone.

\begin{acknowledgments}
The author would like to thank Dr. E. E. Alp for the kind hospitality  of
the Advanced Photon Source at Argonne National Laboratory where this work
was performed.
\end{acknowledgments}


\begin{thebibliography}{}



\bibitem{Seto2000} M. Seto, Y. Kobayashi, S. Kitao, R. Haruki, T.
Mitsui, Y. Yoda, S. Nasu and S. Kikuta, Phys. Rev. B {\bf 61}, 11420
(2000).

\bibitem{Hu2003} M. Y. Hu, W. Sturhahn, T. S. Toellner, P. D. Mannheim, D.
E. Brown, J. Zhao, and E. E. Alp, Phys. Rev. B {\bf 67},
094304 (2003).

\bibitem{Parlinski2004} K. Parlinski, P. T. Jochym, O. Leupold, A. I.
Chumakov, R. $\rm{R\ddot{u}ffer}$, H. Schober, A. Jianu, J.
Dutkiewicz and W. Maziarz, Phys. Rev. B {\bf 70}, 224304 (2004).

\bibitem{Maradudin1965} A. A. Maradudin, Rept. Prog. Phys. {\bf 28}, 331 
(1965).

\bibitem{Mannheim1968} P. D. Mannheim, Phys. Rev. {\bf 165}, 1011 (1968).

\bibitem{Mannheim1971} P. D. Mannheim and S. S. Cohen, Phys. Rev. B 
{\bf 4}, 3748 (1971).


\bibitem{footnote1} In Eq. (\ref{2}) we have used the relation
$\omega_{\rm max}^2=2A_{xx}(0,0)/M$ which holds \cite{Mannheim1968} for
any pure harmonic cubic crystal in the nearest neighbor force-constant
approximation. In Eq. (\ref{3}) we have introduced the pure crystal
lattice Green's function $g_0=g_{xx}(\omega;0,0)$ which we will
describe below. The pure crystal density of
squared frequencies
$\nu(\omega^{\prime 2})=\nu(\omega^{\prime })/2\omega^{\prime}$
which is associated with
$g_0$ is normalized to
$\int_0^{\omega^2_{\rm max}}d\omega^{\prime 2}\nu(\omega^{\prime
2})=1$.

\bibitem{footnote2} The PDOS is referred to as being partial since
 not all the nuclei in a given unit cell are necessarily
$\rm{M\ddot{o}ssbauer}$ active.

\bibitem{footnote3} In general for a perturbation $Q$ the response
function is given by $K(\omega)=\omega {\rm Im}{\rm Tr} GQ$ where the
trace is taken over the relevant $O_h$ representations.

\bibitem{footnote4} The four relations given in Eq. (\ref{23})
respectively follow by setting $(\ell^{\prime},
\alpha,\alpha^{\prime})$ equal to $(0,x,x)$, $(0,x,y)$, $(111,x,x)$ and 
$(111,x,y)$ in Eq. (\ref{13}), restricting the sum on $\ell$ to the
origin and its eight first neighbors, and using the symmetry relations
$A_{xy}(0,0)=0$, $A_{xx}(111)=A_{xx}(\bar{1}11)$,
$A_{xy}(111)=-A_{xy}(\bar{1}11)$, $g_{xy}(0,0)=0$,
$g_{xx}(111)=g_{xx}(\bar{1}11)$,
$g_{xy}(111)=-g_{xy}(\bar{1}11)$, 
$g_{xy}(002)=g_{xy}(020)=g_{xy}(200)=g_{xy}(022)=g_{xy}(202)=0$. In
passing we note that in Ref. \cite{Mannheim1971} the $8A_{xy}(111)g_2=0$ 
relation (a relation which entails that the pure
body-centered cubic lattice Green's function $g_2$ is actually zero in
the nearest neighbor approximation) was not considered as it was not
needed there, with the conclusions reached in the pure central
force-constant study of \cite{Mannheim1968,Mannheim1971} not in any way
being dependent on the value of $g_2$ since the use of the three other 
Green's functions relations given in Eq. (\ref{23}) still allowed for its
elimination in favor of the defect site $g_0$ anyway. However, for the
simplification of the non-central force-constant calculations which are
presented in this paper, the vanishing of $g_2$ turns out to be quite
crucial. 

\bibitem{footnote5} Even if the pure crystal lattice separation is
such that nearest neighbor atoms just happen to be at the same distance
as the minimum in the two body host-host interatomic potential
(central force-constant crystal case), it does not follow that the minimum
in the two body host-defect interatomic potential would occur at
precisely that same distance.

\bibitem{footnote6} In passing we note that there is a difference between
the usage of the real $\omega^2$ equation $1-\rho(\omega^2)S(\omega^2)=0$
of Eq. (\ref{54}) which gives all the modes of the impure crystal and the
usage of the complex $\omega^2$ condition
$1-\rho(\omega^2)S_P(\omega^2)=0$ at which
${\rm Im}G_{xx}(0,0)$ of Eq. (\ref{56}) is resonant, with there being at
best only a handful of frequencies for which this resonant condition is
satisfied, and possibly none at all if $\rho(\omega^2)$ is such that the
$1-\rho(\omega^2)S(\omega^2)=0$ equation has a localized mode solution
outside the pure crystal frequency band. (Outside the band there is no
need to distinguish between $S(\omega^2)$ and $S_P(\omega^2)$ since
$S(\omega^2)$ has no poles there.)  Specifically, in applying Eq.
(\ref{54}) to identify all of the allowed impure crystal eigenmodes
one uses the discrete form of $S(\omega^2)$ given according to Eqs.
(\ref{3}) and (\ref{12}) and Eq. (\ref{57}) below as
$S(\omega^2)=-1-M\omega^2g_{xx}(\omega;0,0)= -(1/N)\sum_{\vec{\bf
k},j}[\sigma_{x}^{*j}(\vec{\bf k})\sigma_{x}^{j}(\vec{\bf
k})\omega_{j}^2(\vec{\bf k})]/[\omega_{j}^2(\vec{\bf k})-\omega^2]$, to
find that the modes of the impure crystal lie either between the modes of
the pure crystal where
$S(\omega^2)$ diverges, or below the lowest pure crystal frequency mode
or beyond the largest pure crystal frequency mode, with the discretized
spectra of the pure and impure crystals having no common eigenfrequency.
(Despite being a sum on both $\vec{\bf k}$ and $j$, the
$\sigma_{x}^{*j}(\vec{\bf k})\sigma_{x}^{j}(\vec{\bf k})$ term in the
expression for $S(\omega^2)$ only permits the $j$ sum to
contribute once, since in the basis in which
$D_{\alpha\beta}(\vec{\bf k})$ is diagonalized we can choose the
polarization vectors $\sigma_{x}^{j}(\vec{\bf k})$ so that only one
polarization eigenvector has a non-zero $x$ component. The function
$S(\omega^2)$ thus only possesses $N$ poles, with the full $3N$
dimensionality of the eigenfrequencies then being recovered via the
additional eigenmode equations for $u_{y}(0,0,0)$ and $u_{z}(0,0,0)$.)
This lack of any common eigenfrequency between the pure and impure
systems holds no matter what values one uses for the $M/M^{\prime}$ and
$A_{xx}(0,0)/A^{\prime}_{xx}(0,0)$ parameters which appear in the
$1-\rho(\omega^2)S(\omega^2)=0$ condition (H. J. Lipkin and P. D.
Mannheim, Bounds on Localized Modes in the Crystal Impurity Problem,
cond-mat/0510542), with the threefold degenerate
$1-\rho(\omega^2)S(\omega^2)=0$ condition always generating a total of
$3N$ perturbed crystal eigenfrequencies. For
perturbations which obey the
$\alpha\hat{\beta} -\beta\hat{\alpha}=0$ condition of Eq. (\ref{52})
then, the threefold degenerate $F_{1u}$ sector condition
$1-\rho(\omega^2)S(\omega^2)=0$ generates the full set of
$3N$ impure crystal eigenmodes, with there being no possibility of any
further modes being generated via eigenvalue conditions associated with
any of the other irreducible $O_h$ representations that are given in Eq.
(\ref{18}).





 
\end{thebibliography}
\end{document}